# PaperBricks: An Alternative to Complete-Story Peer Reviewing*

Jens Dittrich
Saarland University

October 27, 2018

## Abstract

The peer review system as used in several computer science communities has several flaws including long review times, overloaded reviewers, as well as fostering of niche topics. These flaws decrease quality, lower impact, slowdown the innovation process, and lead to frustration of authors, readers, and reviewers. In order to fix this, we propose a new peer review system termed *paper bricks*. Paper bricks has several advantages over the existing system including shorter publications, better competition for new ideas, as well as an accelerated innovation process. Furthermore, paper bricks may be implemented with minimal change to the existing peer review systems.

## 1 Introduction

The current peer review system is heavily criticized in a variety of scientific communities, e.g. [5, 9, 11]. Examples of criticism include from the point of view of the... (1) authors: lack of fairness, intransparency, low quality or superficial reviews, biased reviewers, reviews based on half-read papers, decisions based on one or two reviews only, author feedback with zero impact, overfocus on getting details right, overformalized papers, overselling, frustration — especially for PhD students; (2) readers: flood of syntactically correct yet meaningless papers, delta papers, fostering of niche topics, over-polished papers, suppress of dissent with mainstream ideas, crushing of unpolished yet interesting research ideas and directions, topic killing, missing re-experimentation, no publishing of negative results, biased experimentation, dataset and query picking, long review times, slow innovation process; (3) reviewers: review overload at few times a year, missing reviewing standards and guidelines, huge investment to read a 12-page paper.

This criticism has led to a number of proposals in the past including *open peer review* [6], *post-publication reviews* [11], and double-blind reviewing [8]. However, none of the many proposals has so far been able to substantially heal the many issues of our peer review systems.

In this paper we propose a new system for scientific peer reviews coined *paper bricks*. Paper bricks is very simple to implement as it only requires minimal change to the exsiting peer review system. Yet, paper bricks has many benefits for: (1) authors: less focus on selling details, less focus and paper polishing, less frustration; (2) readers: more high-impact papers, better confidence of experimental results, accelerated innovation process; and (3) reviewers: less investment for reviews, review load spread out over the entire year.

---

*This publication is an extended version of the original presentation shown at the CIDR 2011 Gong Show: The Bowyers. http://www.youtube.com/watch?v=4sorEcLjN04

| Paper brick | Symbol |
|---|---|
| Introduction | I |
| Problem Statement | PS |
| High-Level Solution Idea | HLSI |
| Details | D |
| Performance Evaluation | PE |

Table 1: Possible paper bricks

## 2 Problem Statement

Design a peer review system that solves the above issues. The system should guarantee quality, increase the impact of CS research in industry and other sciences, accelerate the innovation process, and provide a better experience for authors, reviewers, and readers of research publications.

## 3 High-Level Solution Idea

**Observation:** Almost all research papers have the same structure: Introduction, Problem Statement, High-Level Solution Idea, Details, Performance Evaluation.

**Idea:** allow people to publish pre-defined paper sections individually. The possible sections are termed *paper bricks*. Paper bricks may be submitted, reviewed, and published individually. In addition, combinations of paper bricks may be submitted, reviewed, and published as well. Thus, a *publication* consists of one or more paper brick(s). Table 1 shows a list of possible paper bricks. Each publication must be clearly marked to signal the paper bricks it contains on its first page, e.g. this publication is marked with I+PS+HLSI.

We illustrate how this aims at solving the problem statement with a set of examples (Section 3.1 and 3.2). After that we discuss further advantages (Section 3.3) and possible issues (Section 3.4).

### 3.1 Single Paper Brick Publication Examples

**Example 1 (I):** Just publish an Introduction paper brick to draw people's attention to an interesting area, e.g. in the field of bowery an interesting new area might be "Steel Bows". This type of paper brick must be understandable by a broad audience, e.g. an M.Sc. in Bowery should be enough to understand it. The goal of this is to allow people from different communities to link to the same Introduction, e.g. different communities will approach the area with different tools, including possibly different Problem Statements. Introductions with potential interest to a broad community should be selected and republished in magazines such as CACM.

**Example 2 (PS):** Just publish a concise Problem Statement paper brick to crisply define a research problem. This clearly defines the problem to be attacked by the community, e.g. "The Steel Bow Vibration Problem" or "The Steel Bow Arrow Problem". PS-paper bricks must cite at least one I- *or* one PS-paper brick. For each I-paper brick there may be several PS-paper bricks, e.g. in Figure 1 PS-paper bricks 2 and 9 refer to I-paper brick 1. Likewise a PS-paper brick may specialize an existing PS, e.g. in Figure 1 PS-paper brick 19 refer to PS-paper brick 9.

**Example 3 (HLSI):** Just publish a High-Level Solution Idea paper brick to sketch a possible solution. This sketches on a high-level how to solve a particular PS, e.g. "Towards Vibration-free Steel Bows" or "Steel Bows: Copper Arrows to the Rescue?". HLSI-paper bricks must cite at least one PS- or one HLSI-paper brick. For each PS-paper brick there may be several HLSI-paper bricks, e.g. in Figure 1 HLSI-paper bricks 3 and 10 both refer to PS-paper brick 2.

**Example 4 (D):** Just publish a Details paper brick to precisely define the details of an HLSI-paper brick. A D-paper brick usually contains algorithms or detailed descriptions of a method or system, e.g. "How to Calm a Steel Bow" or "Two Copper Arrow Algorithms". `D`-paper bricks must cite at least one `HLSI`- or one `D`-paper brick. For each `HLSI`-paper brick there may be several `D`-paper bricks, e.g. in Figure 1 `D`-paper bricks 21, 22, and 24 all refer to `HLSI`-paper brick 20; in contrast `D`-paper brick 25 refers to `D`-paper brick 24 thus `specializing` paper brick 24, i.e. providing 'details on details'. Notice that the more sideways links you require to reach the root node, the higher the likelihood that the topic being treated is a niche.

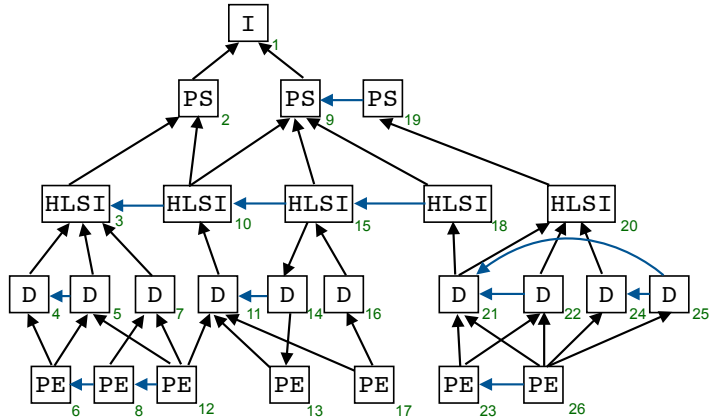

Figure 1: A PaperBrick Graph: Each node is a paper brick. Related Work: **Black** edges are links to paper bricks on a different level. **Blue** edges are links to paper bricks on the same level. **Green** numbers show publication order.

**Example 5** (`PE`): Just publish a Performance Evaluation paper brick to compare one or more algorithms. A `PE`-paper brick presents performance results for one or more `D`-paper bricks, e.g. in Figure 1 `PE`-paper brick 12 refers to `D`-paper bricks 5, 7, and 11. This is what some communities call an "experiments paper". Valid performance evaluation techniques are Measurement, Simulation, and Analytical Modeling [10]. This type of publication allows researchers to validate the performance of different approaches, e.g. "Steel Bows: How Still are they Really?" or "On the Performance of Copper Arrows". Again, for these types of publications it is sometimes only vaguely described what exactly needs to be part of the publication other then experimental results [4], i.e. should the algorithm description be included as well? Is simulation a valid technique? It is [10]!. We believe that only the performance evaluation setup as well as the results and their discussion should be contained in such type of publication. `PE`-paper bricks must cite at least one `D`-paper brick. For each `D`-paper brick there may be several `PE`-paper bricks.

## 3.2 Multiple Paper Brick Publication Examples

**Example 6** (`I+PS`): Publish an Introduction plus Problem Statement. This is interesting for people from industry to make people in academia aware of their problems: "Here is a real problem, please solve it!".

**Example 7** (`I+PS+HLSI` or `I+HLSI`): This is what some communities call a "vision paper". Some conferences have recently started special tracks for vision papers (e.g. PLDI [3], CIDR 2011 [7], or VLDB 2011 [1]). However these publications often do not clearly specify what exactly they should contain; some of these publications do not have a Problem Statement at all. We argue that a "vision paper" should not contain details. In addition, algorithms and experiments are not desired; they may only be used as illustration, i.e. to give directions or to support examples: "This is our vision on what we should be doing."

**Example 8** (`PS+HLSI`): This publication combines a crisp Problem Statement for an existing area with a High-Level Solution Idea. This type of publication allows you to identify new subproblems plus directions in an existing research area.

**Example 9** (`D+PE`): This is basically what we often find in M.Sc. or Ph.D. theses. A `D+PE` publication allows students who initially can neither identify a new problem area, problem statement, nor a high-level solution idea to start with some 'details' for an existing `PS` or `HLSI`. Like that the scope of a student's work is clearly defined without forcing the student to come up with a new `HLSI`-contribution; forcing students to come up with 'something new' often leads to niches. This also allows for **reverse order publications:** a Ph.D.-student may add other paper bricks in reverse order, e.g. he may start his work with a `PE`-paper brick (e.g., paper brick 13 in Figure 1). Then he may come up with ideas for `D`, e.g. new algorithms (paper brick 14), and eventually for a `HLSI` (paper brick 15). In contrast to the existing peer review system the student can easily publish the `PE`-part (assuming an existing implementation) and then later send `D` and `HLSI`.

**Example 10 (`I+PS+HLSI+D+PE`):** The backward compatibility mode. These type of publications are not recommended as they come with all the problems discussed in Section 1.

### 3.3 Advantages

**No early crushing of high-level ideas.** Fancy ideas may be proposed as an `HLSI`-paper brick. Yet, neither details nor algorithms are required. It is fine to have a big market of fancy ideas. People will decide in retrospect which one was the best idea.

**Less investment on selling.** Currently a considerable portion of the paper writing process goes into selling, i.e. justifying the work in the Introduction, contrasting it with other related work, and making sure it is different or has some other twist that was not investigated before. This goes away with paper bricks: it is perfectly fine to have a dozen different `D`-paper bricks for the same `HLSI`-paper brick. None of these works may be rejected with the argument "X considered this problem before". This allows for a better competition. The same holds for `PE`-paper bricks. However it does not hold for `I`, `HLSI`, and `PS`-paper bricks: overlap must be avoided for these paper bricks.

**Less niche topics**. Currently we often see arguments like "Although paper X provided a general solution for problem Y, it did not consider the case where <whatever>. This paper Z fills the gap.". This defense is not required anymore with paper bricks: If there is an `HLSI`- or `D`-paper brick X solving problem Y, it is still fine to write another paper brick Z solving problem Y, if Z differs from X without inventing another niche. Hence, there will be less *forced* niches, yet there will be a big market of solutions.

**Shorter publications.** There should be page limits on paper bricks, e.g. two pages for `I`, one for `PS` and so forth. Writing a shorter publication requires usually less work (unless you prove $P \neq NP$ on a single page). This also allows you to **stream a publication** to a conference: first send `I`. While `I` is being reviewed, you work on `PS` and `HLSI`. Once you receive **early feedback** on `I`, you adjust `PS` and `HLSI`, send a revised `I` plus `PS` and `HLSI`, and so forth. This will help you to avoid wrong directions. It will lower your time spent investing on material that will never make it to a publication. Shorter publications also means less reading, which in turn means **faster review times**. It is also likely that a shorter publication will be reviewed entirely; rather than leaving some 'technical detail in the middle' unreviewed. An additional advantage is that reviewers may specialize for certain types of paper bricks, e.g. some may focus on `HLSI` while others prefer to dive down into `D`.

**Accelerated innovation process.** `I`-, `PS`-, and `HLSI`-paper bricks may be reviewed in days and can then be immediately published online. Thus review times will be considerably shorter as (1) papers bricks are shorter than "full-story"-papers, and (2) the review load is spread out over the entire year (no fixed deadlines whatsoever). Overall, paper bricks creates a big **market of problems, ideas, and details.** That market is easy accessible by both people from industry as well as people from academia at different stages of their career (see Examples 6&9).

**Less Risk.** Assume you have a high-level idea, but are unsure about the details. With the current system there is a high risk for you: if you wait to long to assemble all the pieces for a "full-story"-publication, another researcher may publish your idea before. Then you will not receive *any* credit. With paper bricks this risk does not exist: rather than not publishing your idea at all, you could at least submit an `HLSI`-paper brick early on. If your idea is considered a valuable idea and accepted, you will **receive the credit**. Others may then receive the credit for some `D`-paper brick, but you will receive the credit for `HLSI`.

**Better competition** for best solutions: competition happens at the level of paper bricks rather than "full-story"-publications. In addition, a "full-story"-paper does not *block* (or *kill*) a topic anymore (see Section 3.4).

**Independent experimentation as a principle.** Many fields suffer from publications that present a new algorithm and its performance evaluation at the same time. As the publication pressure is hard and conferences are selective, the experimental results of those publications are often foreseeable: whatever is published is typically considerably better than the previous algorithms. This is also problematic for other researchers who would like to have a good overview on the relative performance of different algorithms. With paper bricks it is easy to write a performance evaluation for any other existing algorithm. There is no special need to justify this or criticize original publications. Repetition of experimental results from a different group is no "strange special case of a publication". It is just fine to send such type of paper brick; in other communities this is good scientific practice anyway. Still, the quality of this `PE`-paper brick must meet certain standards (see Section 3.4).

**Exciting conferences.** Rather than presenting every paper brick that gets accepted, **conferences should select** certain paper bricks for presentation. Presentation slots should be assigned based on relevance to the entire subfield. Paper bricks may be grouped into sessions, e.g. `D`-paper bricks solving the same `PS` should be in the same session. The same holds for `PE`-paper bricks. Notice that Q&A-sessions after a publication presentation are a form of lightweight post-reviewing anyway. As major conferences tend to attract hundreds of domain experts, the collected wisdom of these people should be explored more systematically. Therefore a publication session should be ended with a mini-panel discussing the pros and cons of the different proposals. Panelists should be authors, reviewers, as well as additional domain experts.

Furthermore, conferences may pick up the market idea by having **high-level idea sessions** where the audience votes for the best idea. The Computing Community Consortium [2] already started an initiative in this direction last year [3]. Although this initiative is a step in the right direction, we believe that it is important to clearly define what is part of such type of "vision-paper" and what is not (see Example 7). Again, other than audience voting there should be panel-style discussions allowing for systematic, immediate feedback on the presented ideas — not just three questions or a question by the sessions chair to break the silence. An extension of this could be `HLSI`-**post-reviewing**: only if the vote by the audience is above a certain threshold, the `HLSI`-publication will become permanently visible in the proceedings.

**Connecting the Dots.** Paper bricks form the nodes in a huge graph of research (see Figure 1). Related Work are the edges. Therefore, another idea for conferences could be sessions where a panel plus the auditorium identifying edges, e.g. "this `PS` is also relevant for `I X`!", "this `HLSI` may also solve `PS Y`!", etc.

**Best-of journal publications.** There may still be "full-story" research publications. However, they will look different: In paper bricks a journal publication is created by picking the best paper bricks and republish them as one big publication at a journal. This best-of journal publication identifies the best paper bricks and shows the complete story. For instance, in Figure 1 the path 1, 9, 18, 21, 26 might define a best-of publication.

**Domain Interfacing.** Each paper brick has links to paper bricks on different levels. These links serve as **interfaces to other domains**. For instance, assume you write a `PS`-paper brick in the domain of databases. A domain expert from algorithmic optimization may pick up that `PS` and give hints for a solution in an `HLSI`-paper brick. Details may then be worked out by the database researcher again in a `D`-paper brick. Thus, paper bricks allows people from different domains of CS to collaborate more easily on the same problem.

**Better Assessment of Researchers.** Paper bricks helps students to find a job better matching their qualification after graduating. It is also helpful for job committees in both academia and industry. Which type of paper bricks did she publish? Is she creative in finding high-level ideas? Is she good in details and/or algorithms? Or does she do well in performance evaluations? With the current system it is not always clear who of the authors contributed to which part of a long publication. With paper bricks, this becomes more clear. The specific skills of a researcher become more explicit and thus it becomes easier to assess a student.

| Paper brick | Benchmark |
|---|---|
| Introduction | interesting area? non-existence of similar Introduction in Related Work? understandable by a non-domain expert? etc. |
| Problem Statement | link to some Introduction? correctness? non-existence of similar Problem Statement in Related Work? etc. |
| High-Level Solution Idea | link to some Introduction or Problem Statement? non-existence of similar High-Level Solution Idea in Related Work? etc. Notice: reviewing here could use positive vetoes (cmp. Exciting Conferences). |
| Details | link to some Problem Statement? Link given to other D-paper bricks handling the same Problem Statement? Completeness of the description? Pseudo-code? Runtime Complexity? Space complexity? etc. |
| Performance Evaluation | link to some Details? Clear specification whether Simulation, Experiments, or Analytical Modeling? Meaningful Scenario? Meaningful reference hardware? Meaningful Datasets and Queries? Appropriate Benchmark? Scaling experiments? Appropriate baselines and competitors? etc. |

Table 2: Possible paper brick reviewing guidelines

**Clear contributions.** Most "full-story" research publications make several contributions. With paper bricks, the individual contributions are submitted and reviewed independently. Hence, publications are **easier to judge** by reviewers and readers.

### 3.4 Some Issues and Possible Solutions

**Title, Abstract, and Conclusion** belong to every publication.

**Related Work?** Paper bricks are the nodes; related work are the edges. Each paper brick must provide appropriate edges (see Figure 1 for an example). It is up to the authors to decide whether Related Work is discussed in a separate section or integrated into the main sections.

**How to Review?** Overall: the quality of the write-up matters, i.e. English, brevity, up to the point yet readable. In addition, there should be different standards for the *paper bricks*, see Table 2.

## 4 Conclusions

We have proposed *paper bricks*. *Paper bricks* allows for a fine-granular yet peer-reviewed way of doing research. We believe that paper bricks has strong advantages over the existing peer review systems. As future work we are planning to explore paper bricks in a D-paper brick. We would also like to see conferences trying out our approach. A possible way to bootstrap paper bricks is to add a special *paper bricks track* to an existing conference. If the experiences with such a track is positive, it could be extended to other tracks.

**Acknowledgements.** I would like to thank the attendees from CIDR 2011, numerous colleagues from Saarland University, as well as peers watching my yotube-video for their valuable feedback on the initial idea.